\documentclass[11pt,twoside]{article}

\usepackage{asp2006}
\usepackage{epsf}
\usepackage{psfig}
\usepackage{lscape}

\begin{document}
\title{The XBS sample of type 1 AGNs:
Radio loudness vs. physical parameters}   
\author{A. Caccianiga, P. Severgnini, R. Della Ceca, A. Corral, 
E. Marchese {\it on behalf of the XMM-Newton SSC collaboration}}
\affil{INAF-Osservatorio Astronomico di Brera, Milan, 
Italy}    

\begin{abstract} 
We have studied the relationship between the AGN radio-loudness and the
mass of the central black-hole by using the $\sim$190 type 1 AGN selected
in the XMM-Newton Bright Survey (XBS). We find that radio-loud AGNs are much 
more frequent at high black-hole masses being a factor $\sim$10 times more 
common for masses $>$10$^9$ 
M$_{\sun}$ when compared to masses between 10$^7$-10$^9$ M$_{\sun}$. 
\end{abstract}



\section{The XBS sample of type1 AGN: computing R and M$_{BH}$}
The  XMM-{\it Newton} Bright Survey (XBS, Della Ceca et al. 2004) 
includes an X-ray flux-limited (F[0.5-4.5 keV]$>$7$\times$10$^{-14}$ 
erg s$^{-1}$ cm$^{-2}$) sample which is now almost completely 
identified (92\%, Caccianiga et al. 2008). In the area covered by the
NRAO VLA Sky Survey (NVSS, Condon et al. 1998) there are 186 XBS objects 
that are 
spectroscopically classified as type~1 AGN. 
For all these objects we have computed the radio-loudness parameter (R, 
as in Kellerman et al. 1984) 
using the integrated flux at 1.4 GHz, if detected in the NVSS survey, or
an upper limit on R, if not detected. 
We have then used single-epoch optical spectra to estimate the black-hole 
masses (M$_{BH}$) applying the most recent relationships between M$_{BH}$  
and different 
emission-line widths 
(see Vestergaard \& Peterson, 2006; Vestergaard \& Osmer  2009). 

\section{Results: RL fraction versus black-hole mass}

We have computed the fraction of radio-loud AGN (RL, R$>$10) as a function of 
M$_{BH}$ (Fig. 1, left). We observe a significant 
increase of the fraction of radio-loud (RL) AGN for large M$_{BH}$. 
In particular the RL 
fraction is around 1-4\% for masses of $\sim$10$^8$ M$_{\sun}$ while it 
increases up to 
$\sim$30\% for masses $\sim$5$\times$10$^9$ M$_{\sun}$. 
Since the NVSS flux limit is not deep 
enough 
to guarantee the detection of all the RL AGN present in the XBS, we have 
tested 
the stability of our result by applying the non-parametric method described 
by Avni et al. (1980) to estimate the distribution of the 
missing RL AGN taking into account the upper limits. We find that the 
result discussed above is even more significant when the fraction of 
RL AGN, corrected for the missing objects, is plotted (filled points in 
Fig.1, right panel). 

\section{Conclusions}
We have found that RL AGNs are much more frequent at 
high black-hole masses being a factor $\sim$10 times more common for 
masses $>$10$^9$ 
M$_{\sun}$ when compared to masses between 10$^7$-10$^8$ M$_{\sun}$. 
A similar increase of the RL fraction with M$_{BH}$ has been previously 
suggested by Rafter, 
Crenshaw \& Wiita (2009) using an optically 
selected sample of AGNs.
Recent works (e.g. Sikora, 
Stawarz, Lasota 2007) suggest that the black-hole spin may be the key 
parameter that regulates the radio-loudness of an AGN. Interestingly, N-body 
simulations, aimed at investigating the spin development of the central 
black-hole, have led to the conclusion that more massive black-holes should 
have also higher spin values (e.g. Lagos, Padilla \& Cora 2009). 
Therefore, the observed correlation between the radio-loud fraction and the 
black-hole mass could be explained, at least qualitatively, in the context of 
the so-called ``spin paradigm''. 

\begin{figure}
\plotfiddle{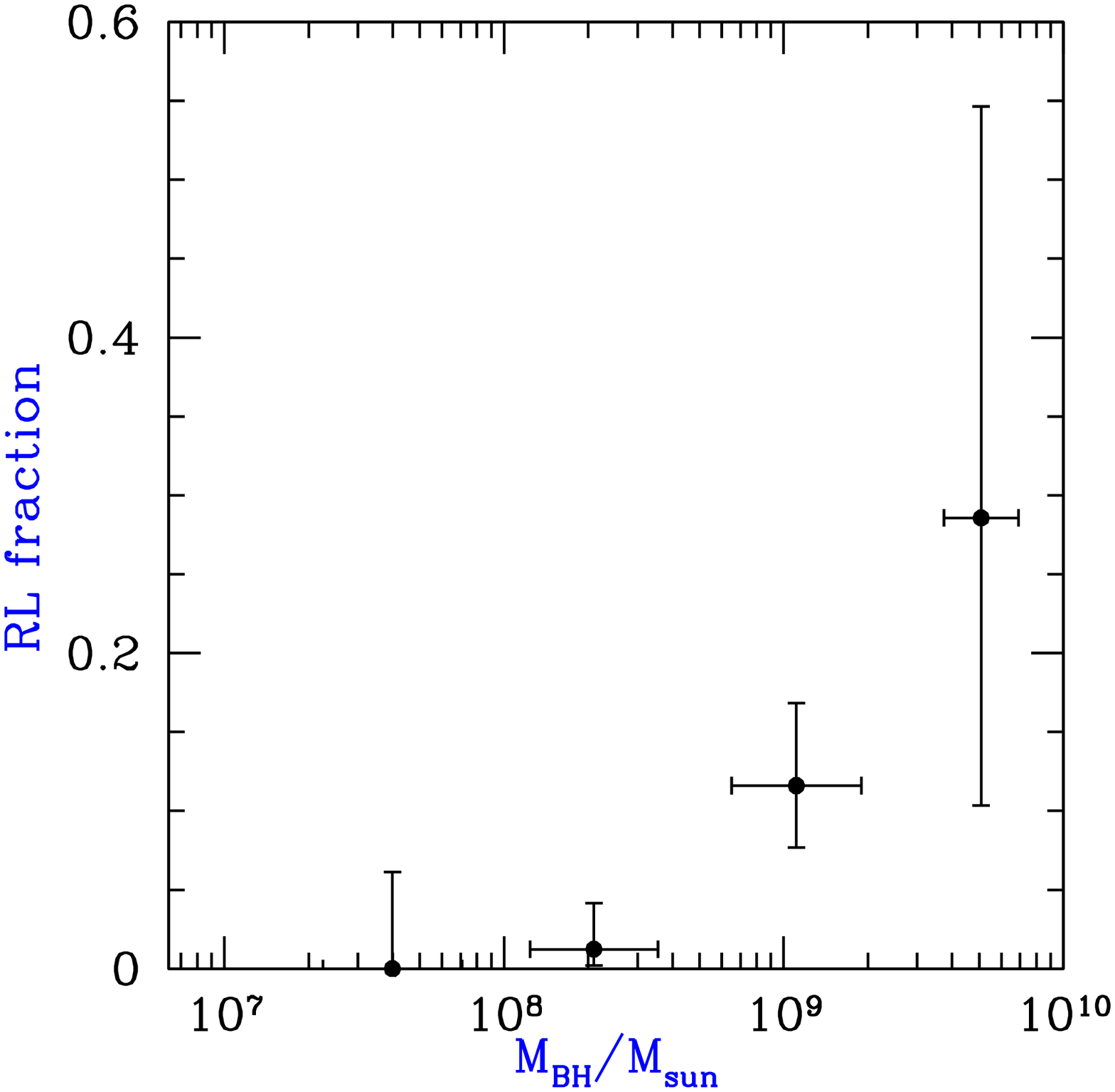}{2cm}{0}{30}{25}{-200}{-115}
\plotfiddle{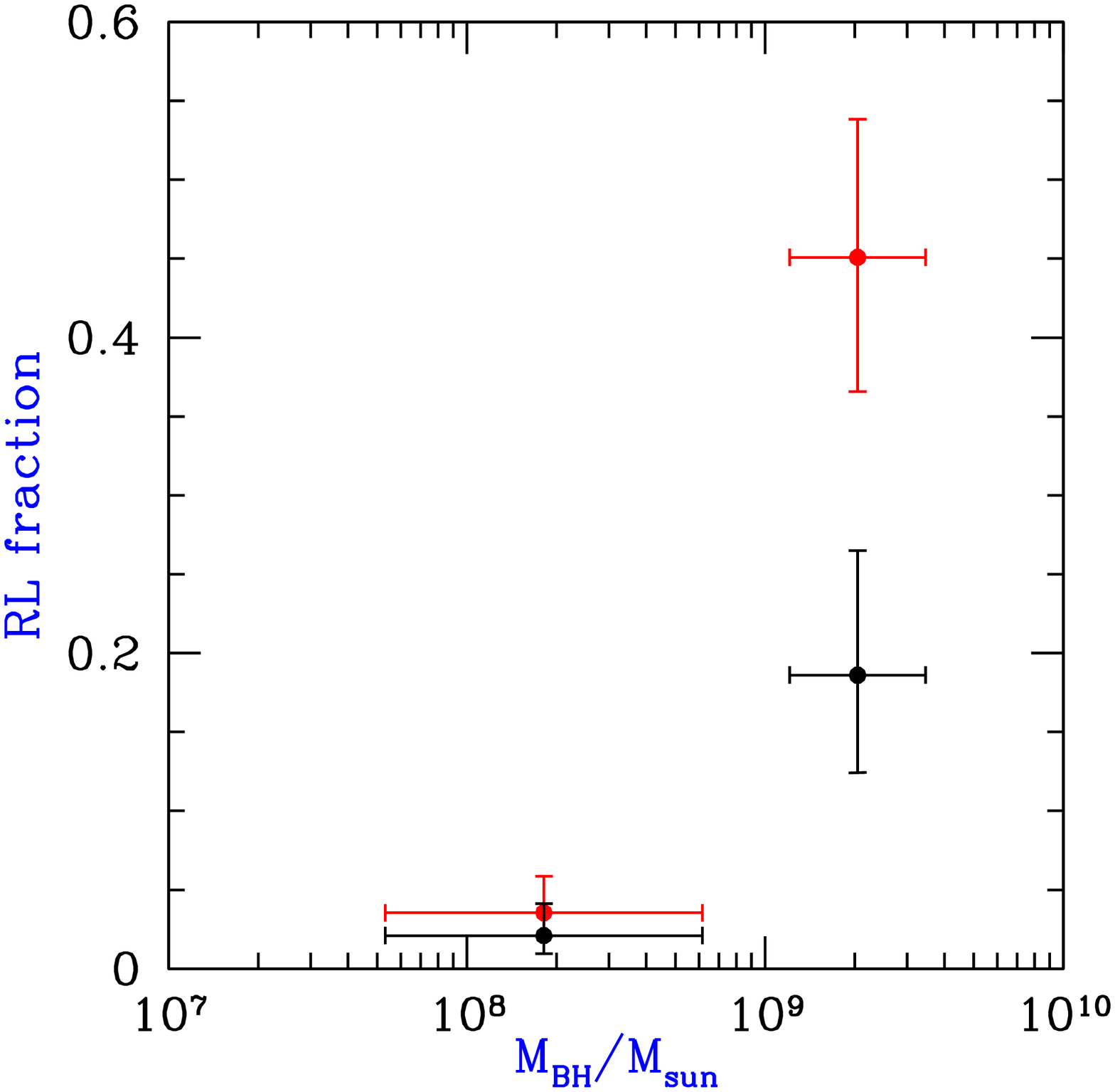}{2cm}{0}{30}{25}{0}{-47}
\caption{{\bf Left}: The fraction of the radio-loud type 1 AGN vs. 
black-hole mass observed in the XBS sample; 
{\bf Right}: The same as in previous figure, with a 
different binning (empty points). Filled points represent the same fraction 
corrected  for the expected number of missing RL AGN.}
\end{figure}
 
\acknowledgements 
We acknowledge financial support from ASI (grant n. I/088/06/0 and COFIS 
contract)
\section*{References}
Avni, Y, Soltan, A., Tananbaum, H., Zamorani, G. 1980, ApJ, 238, 800

\noindent
Caccianiga, A., Severgnini, P., Della Ceca, R. et al. 2008, A\&A, 477, 735

\noindent
Condon, J. J., et al. 1998, AJ, 115, 1693

\noindent
Della Ceca, R., Maccacaro, T., Caccianiga, A. et al. 2004, A\&A, 428, 383

\noindent
Lagos, C., Padilla, N. D. \& Cora, S. A. 2009, MNRAS, 395, 625

\noindent
Rafter, S. E., Crenshaw, D.M., \& Wiita, P. J. 2009, AJ, 137, 42

\noindent
Sikora, M., Stawarz, L. \& Lasota, J.-P. 2007, ApJ, 658, 815

\noindent
Vestergaard, M. \& Peterson, B. M. 2006, ApJ, 641, 689

\noindent
Vestergaard, M. \& Osmer, P. S. 2009, ApJ, 699, 800



\end{document}